\documentclass[conference]{IEEEtran}

\usepackage{amsmath,epsfig,bm,amssymb,amsthm}
\usepackage{psfrag}
\usepackage{cite}
\ifCLASSINFOpdf
\else
\fi

\usepackage{amsmath,epsfig,bm,amssymb,amsthm,balance}

\newcommand{\SRi}{{\mathrm{SR}_i}}

\newcommand{\RDi}{\mathrm{R}_i\mathrm{D}}

\newcommand{\rd}{\mathrm {rd}}
\newcommand{\sr}{\mathrm {sr}}

\newcommand{\s}{\mathbf{s}}
\newcommand{\V}{\mathbf{V}}
\newcommand{\w}{\mathbf{w}}
\newcommand{\y}{\mathbf{y}}
\newcommand{\rb}{\mathbf{r}}
\newcommand{\I}{\mathbf{I}}
\newcommand{\0}{\mathbf{0}}
\newcommand{\A}{\mathbf{A}}
\newcommand{\B}{\mathbf{B}}

\newcommand{\h}{\mathbf{h}}
\newcommand{\x}{\mathbf{x}}

\newcommand{\ub}{\mathbf{u}}
\newcommand{\z}{\mathbf{z}}

\newcommand{\q}{\mathbf{q}}
\newcommand{\Sb}{\mathbf{S}}
\newcommand{\G}{\mathbf{G}}
\newcommand{\X}{\mathbf{X}}
\newcommand{\CN} {\mathcal{CN}}

\newcommand{\diag}{\mathrm {diag}}
\newcommand{\E}{\mathrm {E}}

\newcommand{\Vcb}{\bm{\mathcal{V}}}
\newcommand{\U}{\mathbf{U}}

\newcommand{\C}{\mathbf{C}}

\newcommand{\Sgwb}{\mathbf{\Sigma}_{\ovw}}
\newcommand{\ovy}{\overline{\y}}
\newcommand{\ovS}{\overline{\Sb}}
\newcommand{\ovV}{\overline{\V}}
\newcommand{\ovG}{\overline{\G}}
\newcommand{\ovq}{\overline{\q}}
\newcommand{\ovw}{\overline{\w}}

\newcommand{\ovh}{\overline{\h}}

\newcommand{\Sig}{\mathbf{\Sigma}}
\newcommand\norm[1]{\left\lVert#1\right\rVert}
\newcommand{\bb}{\mathbf{b}}

\IEEEoverridecommandlockouts
\IEEEpubid{\makebox[\columnwidth]{\sf 978-1-4799-2903-0/14/\$31.00~\copyright~2014~IEEE Crown} \hspace{\columnsep}\makebox[\columnwidth]{ }}

\begin{document}
\title{Multiple-Symbol Differential Detection for Distributed Space-Time Coding}

\author{
	\IEEEauthorblockN{M. R. Avendi\IEEEauthorrefmark{1}, Ha H. Nguyen\IEEEauthorrefmark{1} and Nguyen Quoc-Tuan\IEEEauthorrefmark{2}}
\IEEEauthorblockA{\IEEEauthorrefmark{1} University of Saskatchewan, Saskatoon, Canada\\
m.avendi@usask.ca, ha.nguyen@usask.ca }
\IEEEauthorblockA{\IEEEauthorrefmark{2}
Vietnam National University, Hanoi, Vietnam \\
tuannq@vnu.edu.vn
}
}

\maketitle

\begin{abstract}
\label{abs}
Differential distributed space-time coding (D-DSTC) technique has been considered for relay networks to provide both diversity gain and high throughput in the absence of channel state information. Conventional differential detection (CDD) or two-symbol non-coherent detection over slow-fading channels has been examined and shown to suffer 3-4 dB loss when compared to coherent detections. Moreover, it has also been shown that the performance of CDD severely degrades in fast-fading channels and an irreducible error floor exists at high signal-to-noise ratio region. To overcome the error floor experienced with fast-fading, a nearly optimal ``multiple-symbol'' differential detection (MSDD) is developed in this paper. The MSDD algorithm jointly processes a larger window of received signals for detection and significantly improves the performance of D-DSTC in fast-fading channels. The error performance of the MSDD algorithm is illustrated with simulation results under different fading scenarios.
\end{abstract}


\IEEEpeerreviewmaketitle

\section{Introduction}
\label{se:intro}

In distributed space-time coding (DSTC) networks \cite{DSTC-Y}, relays cooperate to combine the received symbols, multiply the results by fixed or variable factors and forward new signals to the destination so that a space-time code can be constructed at the destination. When no channel state information (CSI) is available at the relays and destination, differential DSTC (D-DSTC) scheme has been studied in \cite{D-DSTC-Y} which only needs the second-order statistics of the channels at the relays. Also, the constructed unitary space-time code at the destination provides the opportunity to apply conventional differential detection (CDD) using two received symbols without any CSI. It also has been shown that the performance of the CDD is around 3-4 dB worse than the performance of its coherent version in D-DSTC networks \cite{D-DSTC-Y}.

In practice, the high speed of mobile users leads to time-selective channels. Thus, the common assumption used in differential detection, namely approximate equality of two consecutive channel uses, is violated. Examining the performance of a D-DSTC system shows that the two-symbol differential detection suffers from a severe performance degradation and a high error floor in fast-fading channels.

To overcome the limitations of two-symbol detection in fast-fading channels, in this paper, a near optimal multiple-symbol differential detection for the D-DSTC system is developed. Multiple-symbol differential detection (MSDD), first proposed for point-to-point communications in \cite{msdd-div2}, jointly processes a larger window of the received symbols for detection. As the complexity of MSDD increases exponentially with the window size, the authors in \cite{MSDSD-L} developed a multiple-symbol differential sphere detection (MSDSD) algorithm to reduce the complexity of multiple-symbol detection. Later, reference \cite{MSDUSTC-P} extended the algorithm to unitary space-time codes for MIMO systems. In the context of relay networks, due to the complex form of the distribution of the received signals at the destination, the optimal decision rule of MSDD cannot be easily obtained for the D-DSTC system under consideration. Instead, an alternative decision rule is proposed and the MSDSD algorithm of \cite{MSDUSTC-P} is adapted for the D-DSTC system to provide a detection algorithm with lower complexity. The near optimal performance of the proposed algorithm is illustrated with simulation results in different fading scenarios. It is seen that the proposed MSDD technique, using a window of $N=10$ symbols, is able to significantly improve the performance of the D-DSTC system in fast-fading channels.

The outline of the paper is as follows. Section \ref{sec:system} describes the system model. In Section III, two-symbol differential detection of the D-DSTC system and its performance behavior over time-varying channels are discussed. Section \ref{sec:MSDSD} develops the multiple-symbol differential detection algorithm. Simulation results are given in Section \ref{sec:sim}. Section \ref{sec:con} concludes the paper.

\emph{Notations:} $(\cdot)^t$, $(\cdot)^*$, $(\cdot)^H$, $|\cdot|$ and $\Re\{\cdot\}$ denote transpose, complex conjugate, transpose conjugate, absolute value and real part of a complex vector or matrix, respectively. $\I_R$ and $\0_R$ are $R \times R$ identity matrix and zero matrix, respectively. $\CN(\0,\sigma^2 \I_R)$ stands for circularly symmetric Gaussian random vector with zero mean and covariance $\sigma^2 \I_R$, whereas $\chi_{2R}^2$ is a chi-squared distribution with $2R$ degrees of freedom. $\mbox{E}\{\cdot\}$, $\mbox{Var}\{\cdot\}$ denote expectation and variance operations, respectively. Both ${\mathrm{e}}^{(\cdot)}$ and $\exp(\cdot)$ show the exponential function. $\| \cdot \|$ denotes the Euclidean norm of a vector. $\diag\{x_1,\cdots,x_R\}$ is the diagonal matrix with $x_1,\cdots,x_R$ as its diagonal entries and $\diag\{\X_1,\cdots,\X_N\}$ is $RN\times RN$ block diagonal matrix with the $R \times R$ matrices $\X_l$ on its main diagonal. $\otimes$ is Kronecker product.

\section{System Model}
\label{sec:system}
\begin{figure}[t]
\psfrag {Source} [] [] [1.0] {Source}
\psfrag {Relay1} [] [] [1.0] {Relay 1}
\psfrag {Relay2} [] [] [1.0] {Relay 2}
\psfrag {RelayR} [] [] [1.0] {Relay $R$}
\psfrag {Destination} [] [] [1.0] {Destination}
\psfrag {f1} [r] [] [1.0] {$q_1[k]$}
\psfrag {g1} [l] [] [1.0] {$g_1[k]$}
\psfrag {f2} [bl] [] [1.0] {$q_2[k]$}
\psfrag {g2} [] [] [1.0] {\;\;$g_2[k]$}
\psfrag {fR} [] [] [1.0] {$q_R[k]$\;\;\;}
\psfrag {gR} [l] [] [1.0] {\;\;$g_R[k]$}
\centerline{\epsfig{figure={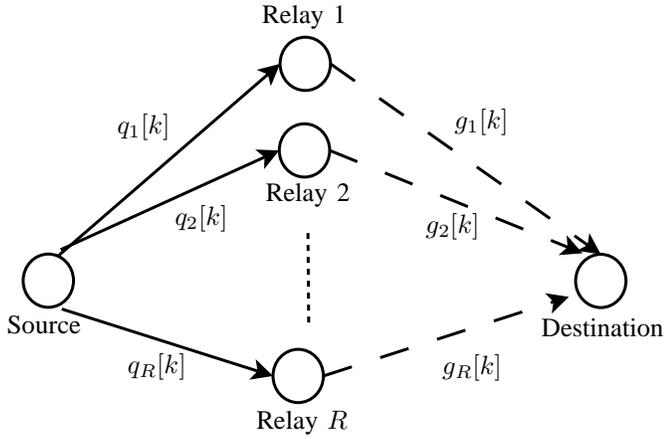},width=8.5cm}}
\caption{The wireless relay network under consideration.}
\label{fig:sysmodel}
\end{figure}

The wireless relay network under consideration, shown in Fig.~\ref{fig:sysmodel}, is the same model as in \cite{D-DSTC-Y}. It has one source, $R$ relays and one destination. Source communicates with Destination via the relays. Each node has a single antenna, and the communication between nodes is half duplex (i.e., each node is able to only send or receive in any given time). Individual channels from Source to the $i$th relay ($\mathrm{SR}_i$) and from the $i$th relay to Destination ($\mathrm{R}_i$D) are Rayleigh flat-fading and spatially uncorrelated.
It is assumed that the variance of all the channels are equal, i.e., channels are symmetric.

Information bits are converted to symbols using a modulation technique such as PSK or QAM at Source. Depending on the number of relays and the type of constellation, appropriate $R\times R$ unitary matrices $\Vcb=\{\V_l| \V^H_l \V_l=\V_l\V^H_l=\I_R,\; l=1,\cdots,L \}$ are used, where $L$ is the total number of codewords. We refer the reader to \cite{D-DSTC-Y} for more details on selecting these matrices.
The transmission process is divided into two phases and sending a codeword (or matrix) from Source to Destination in two phases is referred to as one transmission block indexed by $k$.
 Information symbols are encoded into codeword $\V[k]\in \Vcb$. Before transmission, the codeword is differentially encoded as
\begin{equation}
\label{eq:s[k]}
\s[k]=\V[k] \s[k-1],\quad \s[0]=[1 \quad 0 \quad \cdots \quad 0]^t.
\end{equation}
Obviously, the length of vector $\s[k]$ is $R$.

In phase I, vector $\sqrt{P_0R}\s[k]$ is transmitted by Source to all the relays, where $P_0$ is the average source power per transmission. The transmitted vector is affected by $\SRi,\; i=1,\cdots,R$, channel coefficients which are assumed to be quasi-static during each block but change continuously from block to block. The coefficients of $\SRi,\; i=1,\cdots,R$, channels during the $k$-th block are represented by $q_i[k]\sim \mathcal{CN}(0,1)$. Also, the auto-correlation value between two channel coefficients, which are $n$ blocks apart, follows the Jakes' fading model \cite{microwave-jake}:
\begin{equation}
\label{eq:phi_sr}
\varphi_{\sr}(n)=\E\{q_i[k] q_j^*[k+n]\}=
\left\lbrace
\begin{matrix}
J_0(4 \pi f_{\sr} n R) & i=j \\
0 & i \neq j
\end{matrix} \right.
\end{equation}
where $J_0(\cdot)$ is the zeroth-order Bessel function of the first kind, $f_{\sr}$ is the maximum normalized Doppler frequency of Source-Relay (SR) channels. Also, it is assumed that the carrier frequency is the same for all links.

The received vector at the $i$th relay is
$
\label{eq:ysri}
\rb_i[k]=\sqrt{P_0R}\; q_i[k]\s[k]+\ub_i[k]
$
where $\ub_i[k]\sim \mathcal{CN}(\0,N_0\I_R)$ is the noise vector at the $i$th relay.

The received vector at the $i$th relay is linearly combined with its conjugate as
\begin{equation}
\label{eq:ti}
\x_i[k]=c_i \left(\A_i \rb_i[k]+\B_i \rb_i^*[k]\right)
\end{equation}
where $\A_i$ and $\B_i$ are the combining matrices and determined based on the space-time code that is used for the network. Usually, one matrix is chosen as a unitary matrix and the other one is set to zero. Again, we refer to \cite{DSTC-Y,D-DSTC-Y} for more details on determining the combining matrices. Also, $c_i$ is the amplification factor at the relay that can be either fixed or varying. A variable $c_i$ needs the instantaneous CSI. For D-DSTC, in the absence of CSI, the variance of $\SRi$ channels (here normalized to one) is utilized to define a fixed amplification factor as
$
\label{eq:C_i}
c_i=\sqrt{P_i/(P_0+N_0)}
$, where $P_i$ is the average power per symbol of the $i$th relay. It was shown in \cite{DSTC-Y,D-DSTC-Y} that for a given total power in the network, $P$, for symmetric Source-Relay (SR) and Relay-Destination (RD) channels, $P_0={P}/{2}$ and $P_i={P}/{(2R)}$ form the optimum power allocation between Source and the relays to minimize the pairwise-error probability (PEP). Hence, the amplification factor $c=c_i=\sqrt{{P}/{R(P+2N_0)}},\quad i=1,\cdots,R$, is chosen for all the relays.

Then, in phase II, the relays send their data simultaneously to Destination. Again, using the quasi-static assumption, the coefficients of $\RDi,\quad i=1,\cdots,R$, channel during the $k$-th block are represented by $g_i[k]\sim \mathcal{CN}(0,1)$. Similarly, the auto-correlation value between two channel coefficients, $n$ blocks apart, is expressed as
\begin{equation}
\varphi_{\rd}(n)=\E \{ g_i[k] g_j^*[k+n] \}=
\left\lbrace
\begin{matrix}
J_0(4\pi f_{\rd}nR) & i=j \\
0 & i \neq j
\end{matrix}
\right.
\end{equation}
where $f_{\rd}$ is the maximum normalized Doppler frequency of RD links.

The corresponding received vector at Destination is
\begin{equation}
\label{eq:y}
\y[k]= \sum \limits_{i=1}^{R} g_i[k] \x_i[k]+\z_i[k]
\end{equation}
where $\z_i[k]\sim \mathcal{CN}(\0,N_0\I_R)$ is the noise vector at Destination. Substituting (\ref{eq:ti}) into (\ref{eq:y}) yields \cite{D-DSTC-Y}
\begin{equation}
\label{eq:yii}
\y [k]= c \sqrt{P_0 R} \mathbf{S}[k] \h[k]+\w[k],
\end{equation}
where $\Sb[k],\h[k]$ and $\w[k]$ are the distributed space-time code, the equivalent cascaded channel vector and the equivalent noise vector, respectively, defined as

\begin{equation}
\label{eq:Sk-hk-wk}
\begin{split}
\mathbf{S}[k]&=[\hat{\A}_1 \hat{\s}_1 \; \cdots \; \hat{\A}_R \hat{\s}_R]=\V[k] \mathbf{S}[k-1]\\
\h[k]&=[\; h_1[k]\; \cdots \; h_R[k]\;]^t\\
\w[k]&=c \sum \limits_{i=1}^{R} g_i[k] \hat{\A}_i \hat{\ub}_i[k]+\z_i[k]
\end{split}
\end{equation}
\begin{equation*}
\left.
\begin{aligned}
&\hat{\A}_i=\A_i,\; h_i[k]=q_i[k] g_i[k],\\
&\hat{\ub}_i[k]=\ub_i[k],\; \hat{\s}_i[k]=\s[k]
\end{aligned}
\right\}
\quad  \mbox{if}\quad \B_i=\0
\end{equation*}

\begin{equation*}
\left.
\begin{aligned}
&\hat{\A}_i=\B_i,\; h_i[k]=q_i^*[k] g_i[k],\\
  &\hat{\ub}_i[k]=\ub_i^*[k],\; \hat{\s}_i[k]=\s^*[k]
\end{aligned}
\right\}
\quad \mbox{if}\quad \A_i=\0
\end{equation*}

It should be noted that for given $\{g_i[k]\}_{i=1}^{R}$, $\w[k]$ is $\mathcal{CN}(\0,\sigma_{\w[k]}^2 \I_R)$ where
$
\sigma_{\w[k]}^2=N_0\left(1+c^2 \sum \limits_{i=1}^{R} |g_i[k]|^2\right).
$
It follows that, conditioned on $\Sb[k]$ and $\{g_i[k]\}_{i=1}^{R},$ $\y[k]$ is a circularly symmetric complex Gaussian random vector.

In the following sections, the two-symbol and multiple-symbol differential detections of the received signals at Destination are considered.

\section{Two-Symbol Differential Detection}
\label{sec:CDD}
Coherent detection of transmitted codeword is possible with the knowledge of both SR and RD channels based on the model in \eqref{eq:yii}. In the absence of channel information, in the conventional D-DSTC system that was considered in \cite{D-DSTC-Y}, it is assumed that the channels are fixed for two consecutive block uses. In this case
$
\label{eq:h_slow}
\h[k]\approx \h[k-1]
$
and then from \eqref{eq:s[k]} and \eqref{eq:yii}, one has
\begin{equation}
\label{eq:yk_k-1_slow}
\y[k]=\V[k]\y[k-1]+ \w[k]-\V[k] \w[k-1].
\end{equation}
Given $\y[k]$ and $\y[k-1]$, differential non-coherent detection is applied to detect the transmitted codeword as \cite{D-DSTC-Y}
\begin{equation}
\label{eq:diff_detect}
\hat{\V}[k]= \arg \min \limits_{\V[k] \in \Vcb} \norm{\y[k]-\V[k] \y[k-1] }
\end{equation}
Comparing \eqref{eq:yii} and \eqref{eq:yk_k-1_slow} reveals that the equivalent noise power is enhanced by a factor of two, which explains why the differential non-coherent detection performs approximately $10\log_{10} 2 \approx 3$ dB worse than coherent detection in slow-fading channels.

However, slow-fading assumption requires a coherence interval of $3R$ for both SR and RD channels \cite{D-DSTC-Y} which would be violated for fast-fading channels. Instead, consider that the channel vector $\h[k]$ change by $\Delta \h$  during two consecutive block uses, i.e.,
$
\label{eq:AR-h-hat-vector}
\h[k]=\h[k-1] + \Delta \h.
$
Therefore, one has
\begin{equation}
\label{eq:yk-yk-1}
\y[k]= \V[k] \y[k-1]+\widetilde{\w}[k],
\end{equation}
\begin{equation}
\label{eq:w_tilde}
\begin{split}
\widetilde{\w}[k]=\w[k]-  \V[k]\w[k-1]
+ c\sqrt{P_0R}\; \mathbf{S}[k] \Delta \h.
\end{split}
\end{equation}

As it can be seen, the equivalent noise power is enhanced by an additional factor which is related to the transmitted power and also the amount of channel variation. This means that a higher degradation in the performance of two-symbol deferential detection would be seen in fast-fading channels. Specially, in fast-fading channels, the equivalent noise is dominated by the last term of \eqref{eq:w_tilde}, and as will be seen in Section \ref{sec:sim}, an error floor appears at high signal-to-noise ratio.

\section{Multiple-Symbol Detection}
\label{sec:MSDSD}
As discussed in the previous section, two-symbol differential detection suffers from a large performance degradation in fast-fading channels. To overcome such a limitation, this section develops a multiple-symbol differential detection scheme that takes a window of the received symbols at the destination for detecting the transmitted signals.

Rewrite \eqref{eq:yii} as
\begin{equation}
\label{eq:yii_recall}
\begin{split}
\y [k]&= c \sqrt{P_0 R} \mathbf{S}[k] \h[k]+\w[k]\\
&= c \sqrt{P_0 R} \Sb[k] \G[k] \q[k]+\w[k]
\end{split}
\end{equation}
where $\G[k]=\diag\{g_1[k],\cdots,g_R[k]\}$ and $q[k]=[\; q_1[k],\cdots,q_R[k]\;]^t$.

Let the $N$ received symbols be collected in vector
\begin{equation}
\label{eq:ybar}
\ovy=\left[\; \y^t[1],\y^t[2],\dots, \y^t[N]\; \right]^t,
\end{equation}
which can be written as
\begin{multline}
\label{eq:ovy}
\ovy=c\sqrt{P_0R} \; \ovS \; \ovh +\ovw
=c\sqrt{P_0R}\; \ovS \; \ovG \ovq +\ovw
\end{multline}
$$\ovS= \diag \left\lbrace\; \Sb[1],\cdots, \Sb[N]\; \right\rbrace,$$
$$\ovh=\left[\; \h^t[1],\cdots, \h^t[N]\; \right]^t,$$
$$\ovG=\diag\left\lbrace \; \G[1],\cdots, \G[N] \;\right\rbrace,$$
$$\ovq=\left[\; \q^t[1],\cdots, \q^t[N]\; \right]^t,$$
$$\ovw=\left[\; \w^t[1],\cdots, \w^t[N]\; \right]^t.$$
It should be mentioned that $N$ transmitted symbols collected in unitary block diagonal matrix $\ovS$ correspond to $N-1$ data symbols collected in $\ovV=\diag\{\V[1],\cdots,\V[N-1]\}$ such that
\begin{equation}
\label{eq:SVS}
\Sb[n+1]=\V[n]\Sb[n],\quad n=1,\cdots,N-1
\end{equation}
and $\Sb[N]=\I_R$ is set as the reference symbol\footnote{Note that $\ovS^H\ovS=\ovS\;\ovS^H=\I_{RN}$.}.

Therefore, conditioned on both $\ovV$ (or $\ovS$) and $\overline{\G}$, $\ovy$ is a circularly symmetric complex Gaussian vector with the following pdf:
\begin{equation}
\label{eq:pdfY}
P(\ovy|\ovV,\ovG)=\frac{1}{\pi^N \mathrm{det}\{\Sig_{\ovy}\}} \exp\left( -\ovy^H \Sig_{\ovy}^{-1} \ovy \right).
\end{equation}
In \eqref{eq:pdfY}, matrix $\Sig_{\ovy}$ is the conditional covariance matrix of $\ovy$, defined as
\begin{equation}
\label{eq:RY}
\Sig_{\ovy}=\E \left\lbrace \ovy\; \ovy^H | \ovV,\ovG \right\rbrace=
c^2 P_0R \ovS \; \ovG \Sig_{\ovq} \ovG^H \ovS^H +\Sgwb
\end{equation}
with $\Sig_{\ovq}$ and $\Sig_{\ovw}$ as the covariance matrices of $\ovq$ and $\ovw$, respectively. They are given as follows: 
\begin{equation}
\label{eq:Sigma_qbar}
\Sig_{\ovq}=\E\{ \ovq \;\ovq^H \}=\C_{\ovq}\otimes \I_R,
\end{equation}
$$ \C_{\ovq}=\mathrm{toeplitz}\{ 1,\varphi_{\sr}(1),\dots,\varphi_{\sr}(N-1)\} $$
and
\begin{equation}
\label{eq:Sigma_wbar1}
\Sgwb=\E\{\ovw\; \ovw^H \}=\C_{\ovw} \otimes \I_R
\end{equation}
$$
\C_{\ovw}=
 \diag\left\lbrace \sigma^2_{\w[1]},\cdots,\sigma^2_{\w[N]} \right\rbrace.
$$

Based on \eqref{eq:pdfY}, the maximum likelihood (ML) detection of $N$ transmitted symbols collected in $\ovS$ or the corresponding $N-1$ data symbols collected in $\ovV$ would be given as
\begin{equation}
\label{eq:ML}
\widehat{\ovV}=\arg \max \limits_{\ovV \in \Vcb^{N-1}} \left\lbrace \underset{\ovG}{\E} \left\lbrace
\frac{1}{\pi^N \mathrm{det}\{\Sig_{\ovy}\}} \exp\left( -\ovy^H \Sig_{\ovy}^{-1} \ovy \right)
\right\rbrace \right\rbrace.
\end{equation}
where $\widehat{\ovV}= \diag \left\lbrace\; \widehat{\V}[1],\cdots, \widehat{\V}[N] \; \right\rbrace$.
As it can be seen, the ML metric needs the expectation over the distribution of $\ovG$, which does not yield a closed-form expression. As an alternative, it is proposed to use the following modified decision metric:
\begin{equation}
\label{eq:ML-Modified}
\widehat{\ovV}=\arg \max \limits_{\ovV \in \Vcb^{N-1}} \left\lbrace
\frac{1}{\pi^N \mathrm{det}\{\widehat{\Sig}_{\ovy}\}} \exp\left( -\ovy^H \widehat{\Sig}_{\ovy}^{-1} \ovy \right)
\right\rbrace
\end{equation}
\begin{equation}
\begin{split}
\widehat{\Sig}_{\ovy}=& \underset{\ovG}{\E} \{ \Sig_{\ovy} \} \\ = &c^2P_0 R \ovS  (\C_{\ovh}\otimes \I_R) \ovS^H +(1+c^2 R)N_0 (\I_N \otimes \I_R)\\
=&\ovS \; (\C \otimes \I_R) \; \ovS^H
\end{split}
\end{equation}
\begin{equation}
\label{eq:C}
\C=c^2 P_0 R  \C_{\ovh} +N_0(1+c^2R)\I_N
\end{equation}
\begin{equation}
\label{eq:Sig_h}
\C_{\ovh}=\mathrm{toeplitz} \{ 1,\varphi_{\sr}(1)\varphi_{\rd}(1),\dots,\varphi_{\sr}(N-1)\varphi_{\rd}(N-1) \}.
\end{equation}

Although, the alternative decision metric is not optimal in the ML sense, it will be shown by simulation results that nearly identical performance to that obtained with the optimal metric can be achieved.

Using the rule $\det \{\A\B\}=\det\{\B\A\}$, the determinant in \eqref{eq:ML-Modified} is no longer dependent on $\ovS$ and the modified decision metric can be further simplified as
\begin{equation}
\begin{split}
\label{eq:ML-simp}
\widehat{\ovV}=&\arg \min \limits_{\ovV \in \Vcb^{N-1}} \left\lbrace \ovy^H \widehat{\Sig}_{\ovy}^{-1} \ovy \right\rbrace\\
=& \arg \min \limits_{\ovV \in \Vcb^{N-1}} \{\ovy^H \ovS (\C^{-1}\otimes \I_R) \ovS^H \ovy \}\\
=& \arg \min \limits_{\ovV \in \Vcb^{N-1}} \{\ovy^H \ovS (\U^H \otimes \I_R) (\U \otimes \I_R) \ovS^H \ovy \}\\
=& \arg \min \limits_{\ovV \in \Vcb^{N-1}} \left\lbrace \norm{\bb}^2 \right\rbrace
\end{split}
\end{equation}
where $\U$ is an upper triangular matrix obtained by the Cholesky decomposition of $\C^{-1}=\U^H \U$ and
\begin{equation}
\label{eq:bb}
\bb=(\U \otimes \I_R) \ovS^H \ovy=
\begin{bmatrix}
\sum \limits_{j=1}^{N} u_{1,j} \Sb^H[j] \y[j]\\
\sum \limits_{j=2}^{N} u_{2,j} \Sb^H[j] \y[j]\\
\vdots \\
u_{N,N} \Sb^H[N] \y[N]\\
\end{bmatrix}
\end{equation}
and $u_{i,j}$ is the element of $\U$ in row $i$ and column $j$.

Since $\Sb[N]=\I_R$, the last term of vector $\bb$ does not have any effect on the minimization and it can be ignored. Then by substituting $\Sb^H[n]=\Sb^H[n+1]\V[n]$  (obtained from \eqref{eq:SVS}) into \eqref{eq:bb}, it follows that
\begin{multline}
\label{eq:ML-simp2}
\widehat{\ovV}=\arg \min \limits_{\ovV \in \Vcb^{N-1}} \left\lbrace \sum \limits_{n=1}^{N-1} \| u_{n,n} \V[n] \y[n]  \right. \\  \left. +\Sb[n+1] \sum \limits_{j=n+1}^{N} u_{n,j} \Sb^H[j] \y[j] \| ^2 \right\rbrace
\end{multline}
The simplified alternative minimization in \eqref{eq:ML-simp2} is a sum of $N-1$ non-negative scalar terms
and similar to the decision metric of multiple-symbol detection of unitary space-time coding for MIMO systems given in \cite[eq.5]{MSDUSTC-P}. Therefore, this minimization can be solved using the sphere decoding algorithm described in \cite{MSDUSTC-P} to obtain $N-1$ data symbols with low complexity. 


\section{Numerical Results}
\label{sec:sim}
In this section a relay network with one source, $R=2$ relays and one destination is simulated in different fading scenarios while both two-symbol and multiple-symbol detection schemes are applied. The Alamouti space-time code is chosen for the network. The combining matrices at the relays are designed as \cite{D-DSTC-Y}
$$
\A_1=
\begin{bmatrix}
1 & 0 \\
0 & 1
\end{bmatrix},\;
\B_1=\0, \;
\A_2=\0,\;
\B_2=
\begin{bmatrix}
0 & -1 \\
1 & 0
\end{bmatrix}.
$$
Also, the set of unitary codewords are designed as \cite{D-DSTC-Y}
\begin{equation}
\label{eq:unitary-alamouti}
\mathcal{U}=\left\lbrace \frac{1}{\sqrt{2}}
\begin{bmatrix}
u_1 & -u_2^* \\
u_2 & u_1^*
\end{bmatrix} | u_i \in M-\mbox{PSK}, \quad i=1,2
\right\rbrace.
\end{equation}
The amplification factor at the relays is fixed to $c=\sqrt{{P_i}/{(P_0+N_0)}}$ to normalize the average relay power to $P_i$. The power allocation between Source and the relays is such that $P_0={P}/{2}$ and $P_i={P}/{4},\quad i=1,2$, where $P$ is the total power in the network.

In all simulations, channel coefficients $\{q_i[k]\}_{i=1}^{R}$ and $\{g_i[k]\}_{i=1}^{R}$ are generated independently according to the simulation method of \cite{ch-sim}.
This simulation method has been developed to generate time-correlated fix-to-mobile channel coefficients. The amount of time correlation between the coefficients is determined by the normalized Doppler frequency which is actually a function of the velocity of the mobile user. It is assumed that relays are fixed and then based on the mobility of Source and Destination different cases can be considered. In Case I, it is assumed that the mobility of Source and Destination is low such that all the channels are slow-fading or approximately static and the normalized Doppler frequencies of SR and RD channels are set to $.001$ in the simulation method of \cite{ch-sim}. In Case II, Source and Destination are moving and Source has a slightly higher mobility than Destination such that the normalized Doppler frequency of SR and RD channels are $f_{\sr}=.006,\;f_{\rd}=.004$. In Case III, it is assumed that Source and Destination are moving faster and Destination has a slightly higher mobility than Source such that $f_{\sr}=.009,\;f_{\rd}=.01$. 

\begin{figure}[tb]
\psfrag {Case I} [] [] [.8] {Case I \quad}
\psfrag {Case II} [] [] [.8] {\quad Case II \quad}
\psfrag {Case III} [] [] [.8] {Case III}
\psfrag {P(dB)} [t][] [1]{$P/N_0$ (dB)}
\psfrag {BER} [] [] [1] {BER}
\psfrag {Simulation} [l] [] [.9] {Simulation}
\psfrag {Analysis} [l] [] [.9] {Analysis}
\psfrag {Error floor} [] [] [.8] {Error floors}
\centerline{\epsfig{figure={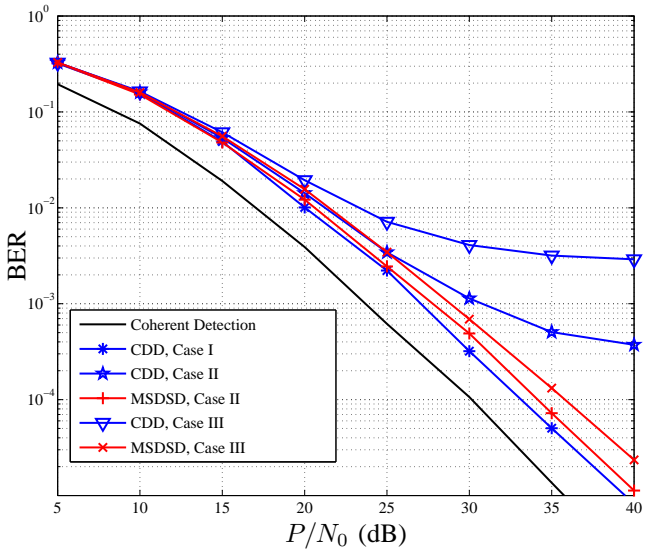},width=8.5cm}}
\caption{ BER of CDD and MSDSD techniques for D-DSTC relaying with two relays in different fading scenarios using Alamouti code and BPSK.}
\label{fig:ber_m2_all}
\end{figure}

\begin{figure}[tb]
\psfrag {P(dB)} [t][] [1]{$P/N_0$ (dB)}
\psfrag {BER} [] [] [1] {BER}
\psfrag {Case I} [] [] [.8] {Case I \quad}
\psfrag {Case II} [] [] [.8] {Case II}
\psfrag {Case III} [] [] [.8] {Case III}
\psfrag {Simulation} [l] [] [.9] {Simulation}
\psfrag {Analysis} [l] [] [.9] {Analysis}
\psfrag {Error floor} [l] [] [.8] {Error floors}
\centerline{\epsfig{figure={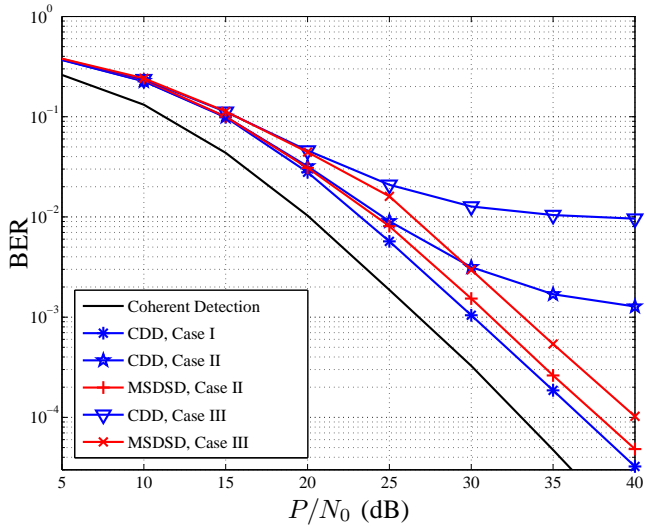},width=8.5cm}}
\caption{ BER of CDD and MSDSD techniques for D-DSTC relaying with two relays in different fading scenarios using Alamouti code and QPSK.}
\label{fig:ber_m4_all}
\end{figure}

To evaluate the BER of the system, in each case, binary data is converted to BPSK/QPSK constellation and then to unitary codewords based on \eqref{eq:unitary-alamouti}. Next, the codewords are encoded differentially according to \eqref{eq:s[k]}. At Destination, two consecutive received codewords are used to estimate the transmitted symbols using the two-symbol differential detection given in \eqref{eq:diff_detect}. The simulation is run for various values of the total power in the network. The practical values of the BER are computed for all cases and plotted versus $P/N_0$ in Figs.~\ref{fig:ber_m2_all}-\ref{fig:ber_m4_all}. For comparison purpose, perfromance of coherent detection of the received symbols for slow-fading channels (Case I) is also evaluated and plotted
in these figures.

As can be seen in Figs.~\ref{fig:ber_m2_all}-\ref{fig:ber_m4_all}, for CDD in Case I, with slow-fading channels, the error probability is monotonically decreasing with $P/N_0$ and the desired cooperative diversity is achieved for the D-DSTC system. Approximately 3-4 dB performance degradation can be seen between coherent and non-coherent detections in this case. However, in CDD Case II, with fairly fast-fading channels, the BER plot gradually deviates from the results in Case I, around $P/N_0=20$ dB, and reaches the error floor after $P/N_0=35$ dB. This phenomenon starts earlier, around $P/N_0=15$ dB, for CDD Case III. The performance degradation is much more severe and error floors of $3 \times 10^{-3}$ (BPSK) and $10^{-2}$ (QPSK) can be seen after $P/N_0=30$ dB.

Given the poor performance of the CDD in Case II and III, MSDSD-DSTC algorithm with $N=10$ is applied to Case II and Case III. Because of the orthogonality of Alamouti space-time code, the decision rule \eqref{eq:ML-simp2} can be divided into two separate decision rules, one for each $u_i, i=1,2$, symbol, and solved separately with low complexity using MSDSD algorithm of point-to-point systems given in \cite{MSDSD-L}. The BER results of MSDSD-DSTC algorithm are also plotted in Figs.~\ref{fig:ber_m2_all} and \ref{fig:ber_m4_all}. Since the best performance is achieved in the slow-fading environment, the BER plot of Case I can be used as a benchmark to see the effectiveness of MSDSD-DSTC algorithm. As can be seen, the MSDSD-DSTC algorithm is able to bring the performance of the system in Case II and Case III very close to that of Case I.

\section{Conclusion}
\label{sec:con}
This paper has shown that using two-symbol detection for differential distributed space-time coding in relay networks  fails to provide a satisfactory performance in fast-fading channels. A near optimal multiple-symbol differential detection algorithm was then developed that can be implemented using sphere decoding with low complexity. Simulation results illustrated that the multiple-symbol detection significantly improves performance of the differential distributed space-time coding systems in fast-fading channels.

\balance
\bibliographystyle{IEEEbib}
\bibliography{references}

\end{document}